\documentclass[twoside,reqno,11pt]{fcaa}

\usepackage{graphicx}
\usepackage{epsfig}
\usepackage{amsthm}
\usepackage{amsmath}
\usepackage{latexsym}
\usepackage{amsfonts}
\usepackage{amssymb}

\newtheoremstyle{theorem}
  {15pt}          
  {15pt}  
  {\sl}  
  {\parindent}
  {\sc}  
  {. }    
  { }    
  {}     
\theoremstyle{theorem}

\newtheoremstyle{defi}
  {15pt}          
  {15pt}  
  {\rm}  
  {\parindent}     
  {\sc}  
  {. }    
  { }    
  {}     
\theoremstyle{defi}

 \def\proofname{\indent\rm P\ r\ o\ o\ f.}
 
 \def\qed{\hfill$\Box$} 

\usepackage{hyperref} 
\usepackage{epstopdf}
\usepackage{amsmath}
\usepackage{amssymb}
\usepackage{dsfont}
\usepackage{trsym}
\usepackage{graphicx}
\usepackage{subfigure}

 \def\theequation{\arabic{section}.\arabic{equation}}

 \def\themycopyrightfootnote{\vspace*{3pt}
 \copyright \, Year\,  Diogenes  Co., Sofia
   \par  \noindent pp. xxx--xxx
   \hfill  \vspace*{-36pt}
  }

\title[Space-fractional diffusion-advection equation]{The space-fractional diffusion-advection equation:
                          Analytical solutions and critical assessment of numerical solutions}
 \author[R.\ Stern, F.\ Effenberger, H.\ Fichtner, T.\ Sch\"afer]{Robin Stern$^{1,2}$, Frederic Effenberger$^{1,3}$, Horst Fichtner$^1$,\\ Tobias Sch\"afer$^4$}

\begin{document}

 \vbox to 2.5cm { \vfill }


 \bigskip \medskip

 \begin{abstract}
The present work provides a critical assessment of numerical solutions of the
space-fractional diffusion-advection equation, which is of high significance
for applications in various natural sciences. In view of the fact that, in 
contrast to the case of normal (Gaussian) diffusion, no standard methods and 
corresponding numerical codes for anomalous diffusion problems have been established
yet, it is of importance to critically assess the accuracy and performance of existing 
approaches. Three numerical methods, namely a finite-diffe\-rence
method, the so-called matrix transfer technique, and a Monte-Carlo method based 
on the solution of stochasic differential equations, are analysed and compared 
by applying them to three selected test problems for which analytical or semi-analytical 
solutions were known or are newly derived. The accuracy and performance differences 
are critically discussed with the result that the use of stochastic differential equations
appears to be advantageous.

 \medskip


 \smallskip


 \end{abstract}

 \maketitle

 \vspace*{-16pt}



 \section{Introduction}\label{sec:1}

\setcounter{section}{1}
\setcounter{equation}{0}\setcounter{theorem}{0}

In recent years the phenomenon of anomalous diffusion has attracted increasing attention, because 
there is growing observational evidence for the presence of such non-classical processes in numerous
natural systems. Corresponding enhanced activities extend to at least three different fields, namely
(i) to the primary natural sciences biology, chemistry, and physics, (ii) to numerical analysis, 
and (iii) to the mathematical foundations.

Regarding the mathematical foundation of anomalous diffusion, Eliazar and Klafter 
\cite{Eliazar-Klafter-2011} have recently reviewed the statistics of sub- and superdiffusion. By
generalizing the Einstein-Smoluchowski diffusion model they provided a unified explanation for the
prevalence of anomalous diffusion statistics. With the latter they emphasized the ubiquity of sub- 
and superdiffusion rather than their `anomalous' nature. 

The progress made with respect to observations and the application of models of anomalous diffusion in 
the mentioned three
natural sciences has been reviewed very recently for each. First, the fact that transport processes in 
biological cells are ubiquitously `anomalous' rather than Gaussian diffusive is by now confirmed with
numerous observations as has been described very recently by Hoefling and Franosch 
\cite{Hoefling-Franosch-2013}, who also discuss the relevant physical models. Second, Volpert et al.\ 
\cite{Volpert-etal-2013} have summarized the knowledge about the front dynamics in anomalous 
diffusion-reaction systems, which are of importance for chemistry of certain systems. And, third, Perrone et al.\ 
\cite{Perrone-etal-2013} give an overview of the nonclassical sub- and superdiffusive plasma transport
in fusion as well as in astrophysical plasmas. 

As for the case of Gaussian diffusion, the solutions of fractional diffusion equations describing 
anomalous diffusion for real-world systems lacking a high degree of both homogeneity and symmetry 
require the application of numerical methods.
However, in difference to the case of Gaussian diffusion leading to a Fokker-Planck equation, generally
accepted standard methods and codes for the numerical solution of fractional Fokker-Planck equations have
not emerged as yet. Nonetheless, a number of approaches have been presented. Perhaps the first publically
available code to solve fractional diffusion-advection equations was that developed by Meerschaert and
Tadjeran in the context of their corresponding studies \cite{Meerschaert-Tadjeran-2004,
Meerschaert-Tadjeran-2006} and \cite{Tadjeran-Meerschaert-2007} for a second-oder accurate method for the 
superdiffusion equation. A brief review of other approaches was recently provided by Sousa 
\cite{Sousa-2012}, who presented a second-order accurate method to solve the fractional 
diffusion-advection equation.

A particular difficulty to be dealt with is that of non-trivial (i.e.\ non-zero and non-homogeneous) 
boundary conditions. Besides the unclear exact physical (or intuitive) meaning of a fractional
derivative (see, e.g.\ \cite{Podlubny-2002}), there is the problem of the non-locality of fractional
derivatives that implies a dependence of the explicit formulation of a fractional differential equation
on the chosen boundary conditions, see, for example, Krepysheva et al.\ \cite{Krepysheva-etal-2006} and 
Garcia-Garcia et al.\ \cite{Garcia-Garcia-etal-2012}.
Another problem is how to treat sub- and superdiffusion simultaneously, as might be necessary in 
anisotropic media or in multi-process systems. The latter can be approached as follows.

Rather than solving a fractional Fokker-Planck equation by methods utilizing discretisations of (partial)
fractional derivatives as mentioned above, one can also resort to a solution of the equivalent system of
(ordinary) stochastic differential equations (SDEs). The necessary generalization of Ito's lemma, stating
the equivalence of a Fokker-Planck equation to a system of SDEs, can be found in the papers by Jumarie \cite{Jumarie-2005} 
as well as by Magdziarz and Weron \cite{Magdziarz-Weron-2007}. They demonstrated that a fractional Fokker-Planck equation is 
equivalent to a system of SDEs with one or more Levy processes instead of Wiener processes.

In view of the various numerical methods, there is an increasing need for critical assessments of these
methods, particularly with respect to the implementation of non-trivial boundary conditions. This can be 
achieved by comparing the results of different methods with analytical solutions and with each other
regarding accuracy as well as performance, respectively. With the present study that is concentrating on space-fractional
differential equations in one space dimension we provide such assessment by first presenting a new analytical solution to the
superdiffusion-advection including a source term. This is preceeded by a presentation of known analytical 
solutions for the case without sources (section 2) and followed by a brief description of related numerical
methods and new extensions of them (section 3), and a comparison of the methods when applied to the same test problems (section 4).
The results are summarized and critically discussed in the concluding section 5.


\section{Analytical solutions to space-fractional diffusion problems}\label{sec:2}
\setcounter{section}{2}
\setcounter{equation}{0}\setcounter{theorem}{0}
For this first assessment of methods we limit ourselves to one-dimen\-sional space-fractional 
diffusion-advection equations. In order to define analytical reference solutions for the 
numerical methods we present in the following two subsections known solutions to the case
without sources as well as a newly derived solution to the case including a source. 

\subsection{Space-fractional diffusion equation without sources}
The one-dimensional space-fractional diffusion equation can be written with the Riesz fractional derivative
$\nabla^{\alpha}$ of order $1<\alpha \le 2$ (see, e.g.\ the monograph by Oldham and Spanier \cite{Oldham-Spanier-1974})
and a constant superdiffusion coefficient $\kappa$ in the form:
\begin{equation}
\frac{\partial f}{\partial t} = \kappa \nabla^{\alpha}f \ ; -\infty < x < \infty \label{FDE} \text{.}
\end{equation}
The Riesz fractional derivative is given by 
\begin{equation}
\displaystyle
\nabla^{\alpha} = - \frac{_{-\infty}\mathrm{D}^{\alpha}_{x} +\,\! _{x}\mathrm{D}^{\alpha}_{\infty}}{2\cos(\!\frac{\pi \alpha}{2}\!)} \text{,}
\label{riesz}
\end{equation}
with the left- and right-sided Riemann-Liouville derivatives in the numerator. 
\\
A standard diffusion problem is defined with the initial condition
\begin{equation*}
f(x,0)  = \delta(x) \nonumber
\end{equation*}
formulated with Dirac's Delta function $\delta(x)$ and the boundary conditions
\begin{equation*}
f(-\infty,t) = f(\infty,t) = 0 \text{.}
\end{equation*}
For the test of numerical against analytical solutions (see below) we consider two representations
of the latter. \\
~\\
\subsubsection{Solution in terms of Fox's $H$-function}
~\\
In this form the solution of \ref{FDE} reads \cite{Jespersen-etal-1999, Mainardi-etal-2005}
\begin{align}
f(x,t) = \frac{1}{\alpha |x|} H_{2,2}^{1,1} \left[\frac{|x|}{(\kappa t)^{1/\alpha}} \bigg| \begin{matrix} (1,1/\alpha),(1, 1/2) \\ (1,1),(1,1/2) \end{matrix}  \right] \text{,} \label{EQ:Greens_FOX}
\end{align}
which can be expressed as a computable series \cite{Metzler-Klafter-2000}. For that purpose we use the abbreviations
\begin{eqnarray*}
C_{j,h} &=& \prod_{j=1,j\neq h}^{m} \Gamma(b_j-B_j(b_h+v)/B_h) \\
D_{j,h} &=& \prod_{j=m+1}^{q} \Gamma(1-b_j+B_j(b_h+v)/B_h) \\
E_{j,h} &=& \prod_{j=1}^{n} \Gamma(1-a_j + A_j(b_h+v)/B_h) \\
F_{j,j} &=& \prod_{j=n+1}^{p} \Gamma(a_j-A_j(b_h+v)/B_h) \text{,}
\end{eqnarray*}
where
\begin{align}
&H_{p,q}^{m,n} \left[z \bigg| \begin{matrix} (a_1,A_1), \dots,(a_p,A_p) \\ (b_1,B_1), \dots,(b_q,B_q) \end{matrix}  \right]= \sum_{h=1}^{m} \sum_{v=0}^{\infty} \frac{C_{j,h}}{D_{j,h}} \frac{E_{j,h}}{F_{j,h}} \frac{(-1)^{v}z^{(b_h+v)/B_h}}{v!B_h} 
\end{align}
and obtain 
\begin{equation}
f(x,t) = \frac{1}{\alpha |x|} \sum_{v=0}^{\infty} \frac{\Gamma(\frac{1}{\alpha}(1+v))}{\Gamma(\frac{1}{2}(1+v))\Gamma(1-\frac{1}{2}(1+v))}\frac{(-1)^v (\frac{|x|}{(\kappa t)^{1/\alpha}})^{1+v}}{v!} \text{.}
\end{equation}
For odd $v$ the denominator yields $\pm \infty$, since $\Gamma(1/2-v/2)=\pm \infty$, thus we carry out the summation over even $v = 2n$, $n\in \mathbb{N}_0$, which leads to
\begin{equation}
f(x,t) = \frac{1}{\alpha (\kappa t)^{1/\alpha}} \sum_{n=0}^{\infty} \frac{\Gamma(\frac{1}{\alpha}(1+2n))}{(2n!)\Gamma(1/2+n)\Gamma(1/2 - n)} \left(\frac{x^{2n}}{(\kappa t)^{2n/\alpha}} \right) \label{Fund_Fox_c} \text{.}
\end{equation}
Expression \ref{Fund_Fox_c} can be simplified by exploiting that 
\begin{equation*}
\Gamma(1/2 + n) = \frac{(2n)!}{n!4^n}\sqrt{\pi}
\end{equation*}
and
\begin{equation*}
\Gamma(1/2 - n) = \frac{n! (-4)^n}{(2n)!} \sqrt{\pi}
\end{equation*}
resulting in
\begin{equation}
f(x,t) = \frac{1}{\alpha \pi (\kappa t)^{1/\alpha}} \sum_{n=0}^{\infty} \frac{\Gamma(\frac{1}{\alpha}(1+2n))}{\Gamma(1+2n)}(-1)^n \left(\frac{x^2}{(\kappa t)^{2/\alpha}} \right)^n \text{.} \label{Fund_Fox}
\end{equation}
Note that for the case $\alpha = 2$ the expected fundamental solution of the (Gaussian) diffusion equation is obtained, since
\begin{align*}
f(x,t) &= \frac{1}{2 \pi (\kappa t)^{1/2}} \sum_{n=0}^{\infty} \frac{\Gamma(\frac{1}{2}(1+2n))}{\Gamma(1+2n)}(-1)^n \left(\frac{x^2}{\kappa t} \right)^n \\
& = \frac{1}{2 \sqrt{\pi \kappa t}} \sum_{n=0}^{\infty} \frac{1}{n!}(-1)^n \left(\frac{x^2}{4 \kappa t} \right)^n\\
& = \frac{1}{2\sqrt{\pi \kappa t}} \exp\left(-\frac{x^2}{4\kappa t} \right) \text{.}
\end{align*}
While equation \ref{Fund_Fox} is the exact solution of the space-fractional diffusion equation its actual computation faces computational difficulties arising from the 
fraction involving the $\Gamma$-function.\\
~\\
\subsubsection{Solution in terms of a Fourier series}
~\\
From an application of the Fourier transformation (see, e.g., Podlubny \cite{Podlubny-1998}) $\mathcal{F} \{ \nabla^{\alpha}h(x)\}= -|k|^{\alpha}\tilde{h}(k) $ it follows 
that $\nabla^{\alpha}\cos(ax) = -a^{\alpha}\cos(ax)$ and, thus, that this transformation can be exploited for space-fractional diffusion equations, too.
In particular, the solution to the given problem \ref{FDE} can be expressed in terms of a Fourier series with $f(x,t) = f(x+2L,t)$, where the initial condition
$f(x,0) = \delta(x)$ is being approximated by the rectangular function 
\begin{align}
f(x,0) = \begin{cases}
1/b, & \text{for } -b/2<x< b/2 \\
0, & \text{for } b/2 < x < L-b/2 \\
-1/b, & \text{for } L-b/2 < x < L + b/2
\end{cases} 
\label{rectangular}
\end{align}
where $b,L\in\mathbb{R}$.
With a sink at $x=L$ the boundary condition $f(x\to \pm \infty,t) =0 $ is fulfilled for $L \to \infty$.
The Fourier series representation of the initial condition \ref{rectangular} is
\begin{equation}
f(x,0) = \sum_{n = 1}^{\infty}\frac{2[1+(-1)^{n+1}]}{n \pi b}\sin\left(\frac{n\pi b}{2L} \right)\cos\left(\frac{n \pi x}{L} \right)
\end{equation}
For the limiting case $b \to 0$, representing the $\delta$-function, it follows:
\begin{equation}
g(x) := \lim_{b\to 0} f(x,0) = \sum_{n=1}^{\infty}\frac{[1+(-1)^{n+1}]}{L}\cos\left(\frac{n\pi x}{L} \right) \text{.}
\label{IC_Fourier}
\end{equation}
Using equation \ref{IC_Fourier} together with the assumption $f(x,t) = g(x)T(t)$ in the space-fractional diffusion equation \ref{FDE} yields
\begin{align}
&\frac{\partial}{\partial t}g(x)T(t) = \kappa \nabla^{\alpha}g(x)T(t) \\
\Rightarrow\;\; &g(x) T'(t) = -\left(\frac{n \pi}{L} \right)^{\alpha} \kappa g(x) T(t) \\
\Rightarrow\;\; &\frac{1}{T} \mathrm{d}T = - \left(\frac{n \pi}{L} \right)^{\alpha} \kappa \mathrm{d}t\\
\Rightarrow\;\; &T(t) = C \exp\left[-\left(\frac{n \pi}{L} \right)^{\alpha} \kappa t \right] \text{, }\;\; C=1\;\; \text{w.l.o.g.}
\end{align}
Hence, the solution to the space fractional diffusion equation can be given in terms of a Fourier series representation
\begin{equation}
f(x,t) = \sum_{n=1}^{\infty}\frac{[1+(-1)^{n+1}]}{L}\cos\left(\frac{n\pi x}{L} \right)\exp\left[-\left(\frac{n \pi}{L} \right)^{\alpha} \kappa t \right] \label{Fund_Four} \text{.}
\end{equation}
For $L\to \infty$ this series representation coincides with the exact solution to equation \ref{FDE} given in \ref{EQ:Greens_FOX}.

\begin{center}
\begin{figure}
\includegraphics[scale=0.9]{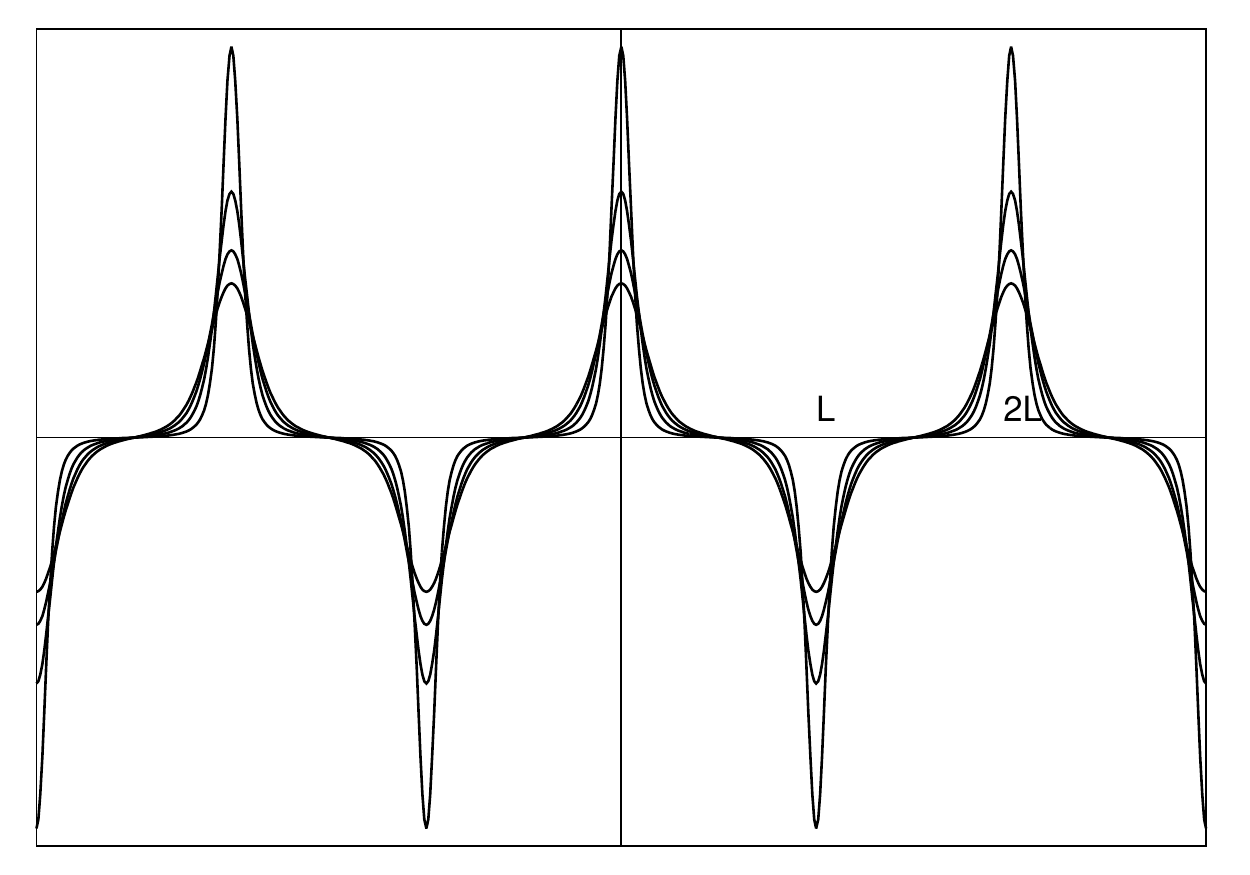}
\caption{\small
Using the Fourier series representation yields a periodic solution to the space-fractional diffusion equation, which converges to the exact solution for $L \to \infty$.
The four lines indicate four different times.
}
\label{Fig1}
\end{figure}
\end{center}


\subsection{Space-fractional diffusion-advection equation with source \label{Problem2}}
The one-dimensional space-fractional diffusion-advection equation with a point source can be written as 
\begin{equation}
\frac{\partial f}{\partial t} = \kappa \nabla^{\alpha}f + a \frac{\partial f}{\partial x} + \delta(x) \label{FDAES}
\end{equation}
where, as above, $\nabla^{\alpha}$ denotes the Riesz fractional derivative, $\kappa$ is a constant superdiffusion coefficient,
and $a$ a constant advection speed.
\\
With the initial condition
\begin{equation}
f(x,0) = \delta(x)
\end{equation}
and the boundary conditions
\begin{equation}
f(-\infty,t) = f(\infty,t) = 0 
\end{equation}
this fractional differential equation can be solved analytically by integrating the fundamental solution to this problem with respect to $t$.\\
While the series representation resulting from Fox's $H$-function (see equation 2.7) cannot be successfully employed due to computational convergence problems for
$t \to 0$, it is possible to integrate the Fourier representation (section 2.2.2), which can be deduced from the previous problem and reads for the case without source as follows
\begin{equation}
f_{\hbox{\tiny no source}}(x,t) = \sum\limits_{n=1}^{\infty} \frac{[1+(-1)^{n+1}]}{L} \cos\left[\frac{n \pi}{L} (x+at) \right] \exp\left[- \left(\frac{n \pi}{L} \right)^{\alpha} \kappa t \right] \text{.}
\label{EQ:Greens_DA}
\end{equation}
Integrating expression \ref{EQ:Greens_DA} with respect to $t$ gives the solution to the space fractional diffusion-advection equation with source:

\begin{align}
 f(x,t) &= \int\limits_{0}^{t} \sum_{n=1}^{\infty} \frac{[1+(-1)^{n+1}]}{L} \cos\left[\frac{n \pi}{L} (x+at') \right] \exp\left[-\left(\frac{n \pi}{L} \right)^{\alpha} \kappa t' \right] \mathrm{d}t' \nonumber\\
& =\sum_{n=1}^{\infty}\Biggl\{[1+(-1)^{n+1}] \biggl[ \left(\frac{n \pi}{L}\right)^{\alpha} \kappa L \cos\left(\frac{n \pi x}{L} \right) -n \pi a \sin\left(\frac{n \pi}{L}x \right) \nonumber \\
& -\left(\frac{n \pi}{L} \right)^{\alpha} \kappa L \cos\left[\frac{n \pi}{L}(x+at) \right] \exp\left[-\left(\frac{n \pi}{L} \right)^{\alpha} \kappa t \right] \nonumber \\
& + n \pi a \sin\left[ \frac{n \pi}{L} (x+at) \right] \exp\left[-\left(\frac{n \pi}{L} \right)^{\alpha} \kappa t \right] \biggr] \text{\Huge{/}} \nonumber \\
& \left[  \left(\frac{n \pi}{L}\right)^{2 \alpha}\kappa^2L^2 + n^2 \pi^2a^2  \right] \Biggr\} \label{SOL:FDAES}
\end{align}
As a result of the periodicity, the solution $f(x, t)$ does not converge to the exact solution for $t \to \infty$, since, in the course of time, the advection $a$ leads to an overlap of the upwind and downwind interval. Thus, for application purposes $L$ has to be chosen properly with regard to the time interval of interest.
This proves to result in a very good approximation to the solution of equation \ref{FDAES} that does not suffer
from convergence problems (see sections 4.1 and 4.2). Furthermore, with a glance at the actual computation, a
user might take advantage from the parallelisation capability of the series representation.
To the best of our knowledge this representation of the solution of equation \ref{FDAES} has not been published
before.


\smallskip
\section{Numerical solution methods for space-fractional diffusion problems}\label{sec:3}
We discuss three methods to solve space-fractional differential equations numerically. 
The first two belong to the class of grid-based schemes, the third is a Monte-Carlo
approach based on stochastic differential equations. 

\setcounter{section}{3}
\setcounter{equation}{0}\setcounter{theorem}{0}

\subsection{Grid-based Schemes}
\subsubsection{A Gr\"unwald-Letnikov-based finite-difference approach by Meer\-schaert and Tadjeran}
~\\
Meerschaert and Tadjeran \cite{Meerschaert-Tadjeran-2006}
presented a finite-difference approximation for space-fractional partial differential equations, which we summarize and extend here for the case of the fractional
Fokker-Planck equation.
Starting from the space-fractional diffusion equation given in (\ref{FDE}) the fractional derivatives are expressed in terms of the so-called Gr\"unwald-Letnikov representation
\cite{Podlubny-1998}
\begin{align*}
\frac{\partial f(x,t)}{\partial t} =& \kappa_+(x,t)\lim_{N \to \infty} \frac{1}{h_+^{\alpha}}\sum_{j=0}^{N}\frac{\Gamma(j-\alpha)}{\Gamma(-\alpha)\Gamma(j+1)} f(x-jh_+,t)\\
+& \kappa_-(x,t) \lim_{N \to \infty}\frac{1}{h_-^{\alpha}}\sum_{j=0}^{N}\frac{\Gamma(j-\alpha)}{\Gamma(-\alpha)\Gamma(j+1)} f(x+jh_-,t) +q(x,t) \text{,} \nonumber
\end{align*}
where $h_+ = (x-L)/N$ and $h_- = (R-x)/N$ with $N$ the number of steps, and $x \in [L,R]$. The superdiffusion coefficients $\kappa_\pm$ might, in general, depend on $x$ and $t$. The discretization considering $1<\alpha\le2$ reads 
for the case $\kappa_+(x,t)=\kappa_-(x,t)=\kappa = const$, which is of interest here (see equation \ref{FDAES}), 
as follows
\begin{equation}
\frac{f_i^{n+1}-f_i^{n}}{\Delta t} = \frac{\kappa}{h^{\alpha}}\biggl(\sum_{j=0}^{i+1}g_j f_{i-j+1}^{n+1}+ \sum_{j=0}^{N-i+1}g_j f_{i+j-1}^{n+1} \biggr)+q_i^{n+1} \text{,} \nonumber
\end{equation}
where $h = (R-L)/N$, $f_i^{n}=f(x(i),t(n))$ and $g_j = \Gamma(j-\alpha)/[\Gamma(-\alpha)\Gamma(j+1)]$. This implicit Euler method can be extended by including an advection term $a \partial f/\partial x$ in order to solve particular cases (here $a=const$) of the space-fractional Fokker-Planck equation: 
\begin{align}
\frac{f_i^{n+1}-f_i^{n}}{\Delta t} = &\frac{\kappa}{h^{\alpha}}\biggl(\sum_{j=0}^{i+1}g_j f_{i-j+1}^{n+1}+ \sum_{j=0}^{K-i+1}g_j f_{i+j-1}^{n+1} \biggr) \nonumber\\
&- \frac{a}{2 h}\left(f_{i+1}^{n+1} - f_{i-1}^{n+1} \right) + q_i^{n+1} \text{.}
\end{align}
This way one obtains a system of linear equations
\begin{align}
{\bf{f}}^{n+1} &= \underline{A} {\bf{f}}^{n+1} + \underline{B} {\bf{f}}^{n+1} + {\bf{f}}^{n} + \Delta t {\bf{q}}^{n+1} \nonumber \\
\underbrace{\left[\underline{I} - \left(\underline{A} +\underline{B} \right) \right]}_{= \underline{M}} {\bf{f}}^{n+1} &= \Delta t {\bf{q}}^{n+1} + {\bf{f}}^{n} \nonumber \\
 {\bf{f}}^{n+1} &= \underline{M}^{-1} \left(\Delta t {\bf{q}}^{n+1} + {\bf{f}}^{n} \right) \label{NUM:1}
\end{align}
The resulting matrix $\underline{A}$ corresponding to space-fractional diffusion reads
\begin{align*}
\begin{small}
\underline{A} = 
\begin{pmatrix}  
0 & 0 & ... & 0 & 0 & 0\\
 \sigma (g_2 + g_0) & 2 \sigma g_1 &  ... &\sigma g_{K-2} & \sigma g_{K-1} & \sigma g_K \\
\sigma g_3 & \sigma (g_2 + g_0) & ...&\sigma g_{K-3} &\sigma g_{K-2} &\sigma g_{K-1} \\
\vdots & \ddots & \ddots &  \ddots & \ddots & \vdots \\
\sigma g_{K-1} & \sigma g_{K-2} & \cdots  & \sigma (g_2 + g_0) & 2 \sigma g_1 & \sigma (g_0 + g_2)  \\
0 & 0 & \cdots & 0 & 0 &0
\end{pmatrix} \text{,}
\end{small}
\end{align*}
where $\sigma = {\kappa\Delta t}/{h^{\alpha}}$, 
while the matrix corresponding to advection is
\begin{align*}
\begin{small}
\underline B = \frac{-a \Delta t}{2h}
\begin{pmatrix}
1 & 0 & -1 & 0 & \cdots & 0 \\
0 & 1 & 0 & -1 & \cdots & 0 \\
\vdots & \ddots & \ddots & \ddots & \cdots & 0 \\
0 & \cdots & \cdots & 1 & 0 & -1
\end{pmatrix} \text{.}
\end{small}
\end{align*}
The solution of this system of linear equations requires an initial condition $f(x,0)$ and boundary conditions $f(L,t)$ and $f(R,t)$. Inverting the matrix $\underline{M}$ can be done, for example, by Gaussian elimination.
According to Meerschaert and Tadjeran \cite{Meerschaert-Tadjeran-2006} this method is unconditionally stable and of the order $O(\Delta t) + O(\Delta x)$.
~\\
\subsubsection{The Matrix Transfer Technique by Ilic et al.}
~\\
An alternative numerical method to solve the space-fractional diffusion equation was presented by Ilic et al.\
\cite{Ilic-etal-2005, Ilic-etal-2006}. These authors considered the standard approximation of the normal diffusion 
equation leading to the expressions
\begin{align}
\frac{\partial f}{\partial t} = -\frac{\kappa}{h^2}(-f_{i+1} + 2f_{i} - f_{i-1})\\
f_0 = g_1(t), \ f_N = g_2(t)\\
f(x,0) = u(x) \text{,}
\end{align}
where $h$ is the step size in space, which can be expressed as a system of ordinary differential equations
\begin{equation}
\frac{d \bf{\Phi}}{d t} = -\eta {\bf{A \Phi}} + \bf{b} \text{.}
\end{equation}
Here, $\bf{A}$ denotes the usual tridiagonal matrix resulting from the discretization of the Laplace operator.

For $\bf{b} = 0$, 
i.e.\ homogeneous boundary conditions, Ilic et al.\ found that the space-fractional diffusion equation can be approximated by
\begin{equation}
\frac{d \bf{\Phi}}{d t} = - \tilde{\eta} {\bf{A}}^{\alpha/2} {\bf{\Phi}} \text{,}
\end{equation}
where $\tilde{\eta} = {\kappa}/{h^{\alpha}}$ and $\alpha$ is the order of the fractional derivative. For further information on this numerical method the reader is referred to \cite{Ilic-etal-2005, Ilic-etal-2006}.\\
For fractional differential equations involving an advection and source term, such as
\begin{equation}
\frac{\partial f}{\partial t} = \kappa \nabla^{\alpha}f + a\frac{\partial f}{\partial x} + q(x) \text{,}
\end{equation}
this method can be extended. Analogously to the procedure in \cite{Ilic-etal-2005,Ilic-etal-2006}, we obtain the equation
\begin{equation}
\frac{d \bf{\Phi}}{d t} = {\bf{\tilde{A}} \Phi} + \bf{Q} \text{,}
\end{equation}
with ${\bf\tilde{A}} = \tilde{\eta} {\bf{A}}^{\alpha/2} + a {\bf{C}}$. $\bf{C}$ is the discrete representation of the advection term.
This equation has the solution
\begin{align}
{\bf{\Phi}}(t) &= e^{{\bf{\tilde{A}}}(t-t_0)} {\bf{\Phi}}_0 + e^{{\bf{\tilde{A}}}t}  \int_{t_0}^{t} e^{{\bf{\tilde{A}}}s} {\bf{Q}} \mathrm{d}s \\
               &=  e^{{\bf{\tilde{A}}}t} \left({\bf{\Phi}}_0 +  {\bf{\tilde{A}}}^{-1} \left( {\bf{I}} - e^{ - {\bf{\tilde{A}}}t}\right) {\bf{Q}} \right)  \text{.}
\end{align}
The second equality holds if (w.l.o.g.) $t_0=0$ is chosen. 


\subsection{Monte-Carlo Method}

To solve a space-fractional diffusion-advection equation like
equation (\ref{FDAES}), it is also possible to recast the problem
  into a stochastic differential equation (SDE), which is then solved
  for a large number of trajectories of pseudo-particles to yield the solution 
  by an appropriate average over the positions of the latter. For standard diffusion, this is a
  well-known approach, see e.g.\ Kopp et al.\ \cite{Kopp-etal-2012} for a discussion
  of the numerical requirements and details, which are partly applied
  in this work as well. The extension to the superdiffusive case is a
  more recent development, and the basic idea and some numerical aspects
  are discussed in \cite{Magdziarz-Weron-2007}. There, also the case of
  competing super- and subdiffusion is included, while we restrict
  ourselves here to pure superdiffusion. Consequently, the SDE for the
  stochastic process $X(t)$ considered is of the general form:
  \begin{equation}
    \label{eq:sde}
    dX(t) = -a(X(t))dt + \kappa^{1/\mu}dL_\mu(t)
  \end{equation}
  where $L_\mu(t)$ is a symmetric $\mu$-stable L\'evy motion, which
  has the Fourier transform property
  \begin{equation}
    \label{eq:levycharact}
    \mathcal{F}\{e^{ikL_{\mu}(t)}\} = e^{-t|k|^{\mu}} \,.
  \end{equation}
  Numerically, this L\'evy process is represented by drawing the
  pseudo-random numbers from the respective distribution, which can be
  calculated for example with the Chambers-Mallows-Stuck method
  \cite{Chambers-etal-1976}. In the case of constant advection and superdiffusion 
  coefficients $a$ and $\kappa$ in the SDE, the pseudo-particles can be traced forward
  or backward in time, resulting in the same solutions to the corresponding
  diffusion-advection equation. For the discussion on how to treat
  initial conditions, sources and boundaries with the SDE approach,
  the reader is referred again to \cite{Kopp-etal-2012}. 

\section{Comparisons of numerical approximations with analytical solutions}\label{sec:4}

\setcounter{section}{4}
\setcounter{equation}{0}\setcounter{theorem}{0}

In the following we discuss the accuracy and performance of the three numerical methods presented in section~3,
namely the finite-difference me\-thod developed by Meerschaert and Tadjeran \cite{Meerschaert-Tadjeran-2006}, the
matrix transfer technique introduced by Ilic et al.\ \cite{Ilic-etal-2005,Ilic-etal-2006}, and the SDE-based Monte-Carlo 
method by Magdziarz and Weron \cite{Magdziarz-Weron-2007}, to three selected test 
problems for which one can obtain analytical solutions as described in section~2.

\subsection{Space-fractional diffusion equation}
\begin{figure}
\begin{center}
\includegraphics[scale=0.8]{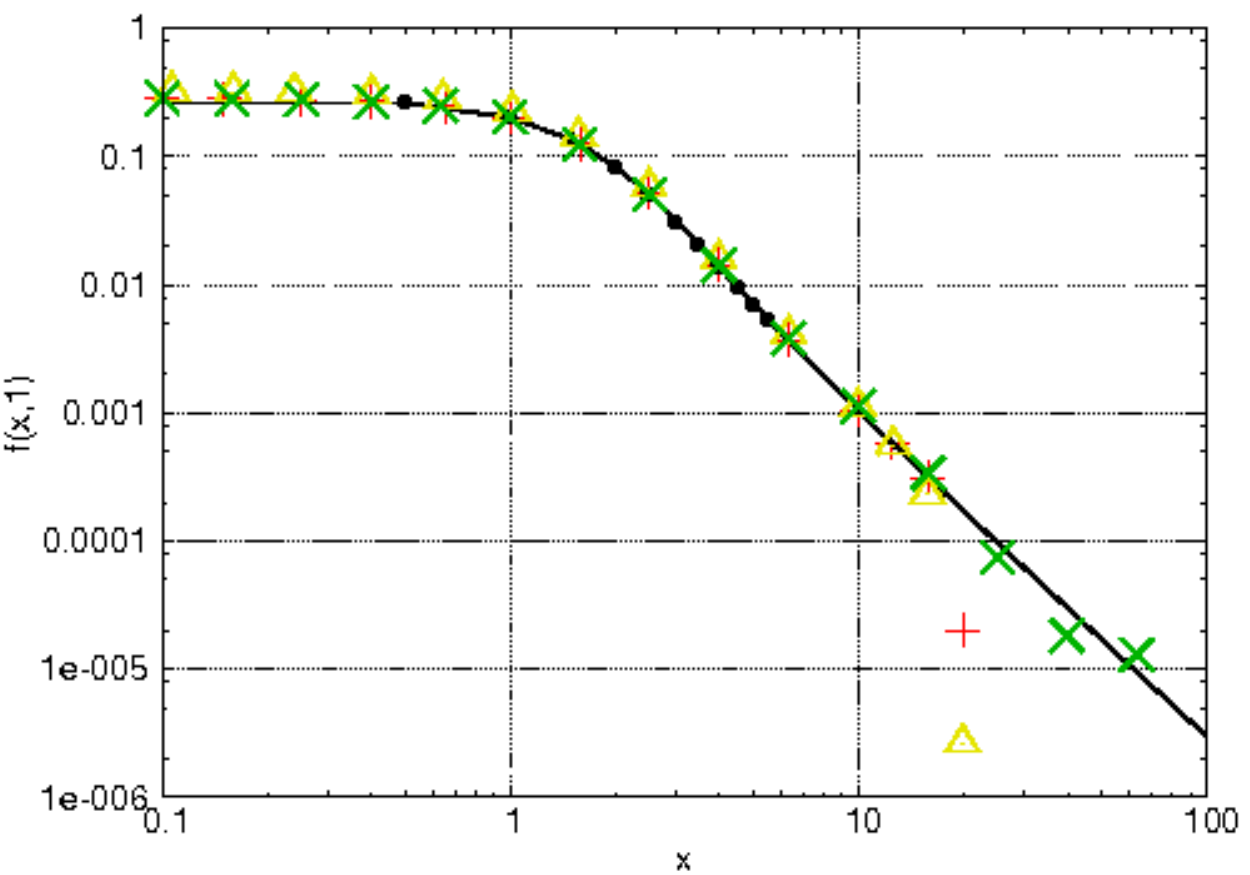}
\caption{\small Numerical and analytical solution to the fractional diffusion equation \ref{FDE} with $\kappa=1$ and $\alpha=3/2$ at $t=1$.
                Black solid line: Fourier series solution \ref{Fund_Four} ($N_{max} = 10^5, L = 10^3$). 
                Black dots: Power series \ref{EQ:Greens_FOX} ($N_{max} = 190$), note the convergence problem for $x > 5.7$, see text.
                Yellow triangles: Matrix transfer technique ($\Delta x \approx 0.03$).
                Red crosses: Finite-difference method ($\Delta x = 0.05, \Delta t = 0.01$).
                Green crosses: SDE ($n_{particles} = 10^6, \Delta t = 0.001$).}
\label{FIG:PROPAGATOR}
\end{center}
\end{figure}

At first we compare the numerical approximations to the exact solution of the space-fractional diffusion equation
\ref{FDE} with $\kappa=1$ and $\alpha=3/2$.
The result is presented in Figure \ref{FIG:PROPAGATOR}.
\vspace*{0.2cm}\\

\noindent
$\bullet$ {\it Analytical solutions:}\\
The representation of the analytical solution using Fox's $H$-function \ref{Fund_Fox} and that given by the Fourier series \ref{Fund_Four} are in good agreement in the interval from $x \in [0,5.7]$. In this particular example the power series cannot be evaluated for $x>5.7$ due to computational accuracy issues related to the evaluation 
of the $\Gamma$-function. In contrast, the Fourier series solution can be evaluated for an arbitrary $x$, demonstrating the power of this representation compared to that
given with \ref{Fund_Fox}. 
A serious drawback of the Fourier series solution is, however, the periodicity of the approximation of the $\delta$-function. For large $x$ and $t$ one has to make sure that
the period $2L$ has been chosen sufficiently large to avoid an interference with the sink at $L$ (see section 2.1.2).
\vspace*{0.2cm}\\
\noindent
$\bullet$ {\it Numerical solutions:}\\
The numerical solutions prove to be a good approximation to the exact solution up to about $x=20$, where the boundary conditions $f(-20,t) = f(20,t) = 0$ 
were chosen (to limit the required computational time). The latter start to corrupt the solution for the unbounded interval beyond $x=15$.  This `power-law decay' is
problematic for grid-based methods, whose accuracy can, however, be improved by extending the $x$-range. This, of course, is on the expense of 
computational time. \\
In contrast to these grid-based numerical methods, the SDE solution is not influenced by any boundary
conditions, since there is no particular finite computational domain (like the grid) on which the solution
has to be determined. Instead, only the number of particles limits the accuracy and reproduction of the 
expected power law, which can thus easily extend over a much wider range.

\subsection{Space-fractional diffusion-advection equation}
\begin{figure}
\subfigure[All methods produce the same shape of the solution. The matrix transfer technique result is corrected by a factor of about 0.03 to fit the other solutions, see
text.]
{
\includegraphics[scale=0.8]{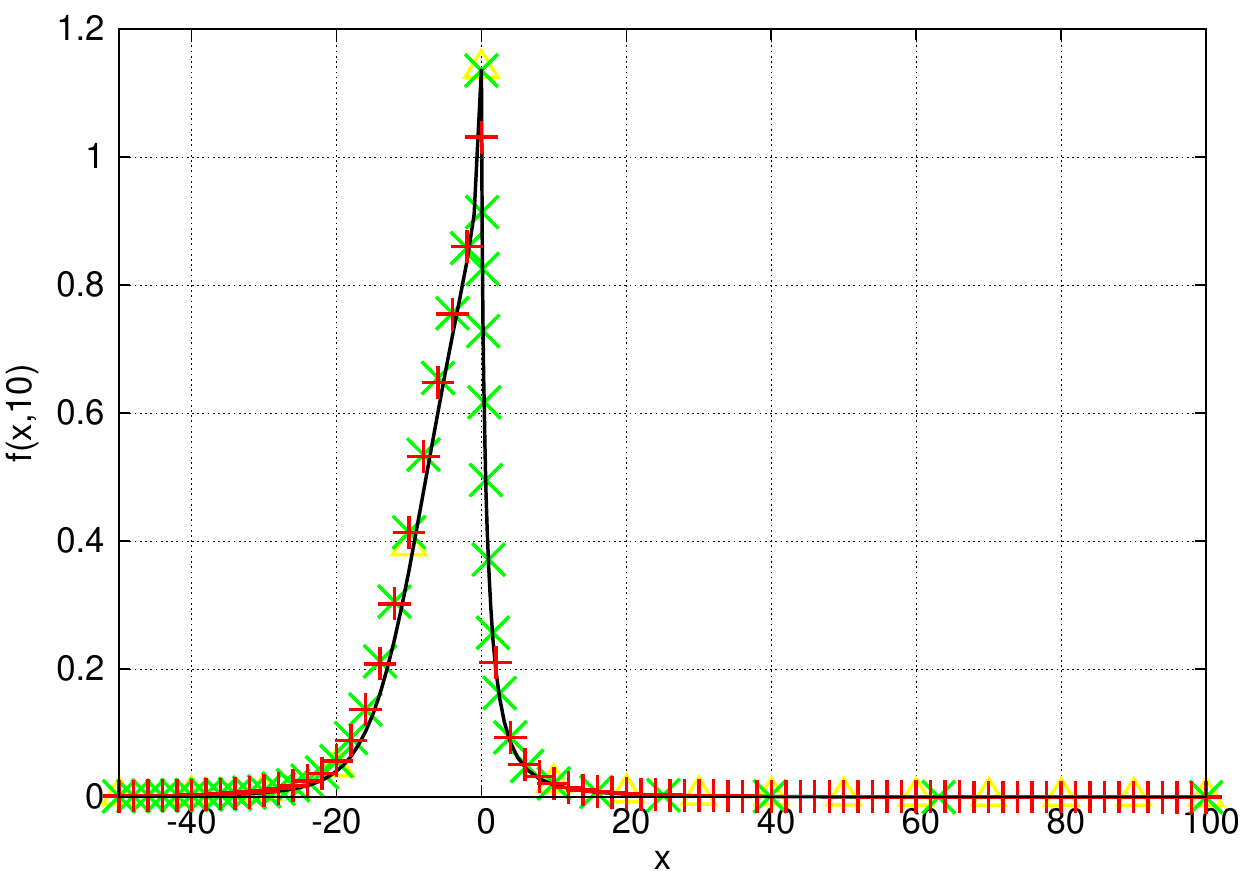}
\label{FIG:FDAES}
}
\subfigure[The logarithmic scaling reveals, that all methods produce the same power-law behaviour. The grid-based methods have boundary values of 0 at $x = 100$ and 
$x = 1000$, respectively.]
{
\includegraphics[scale=0.8]{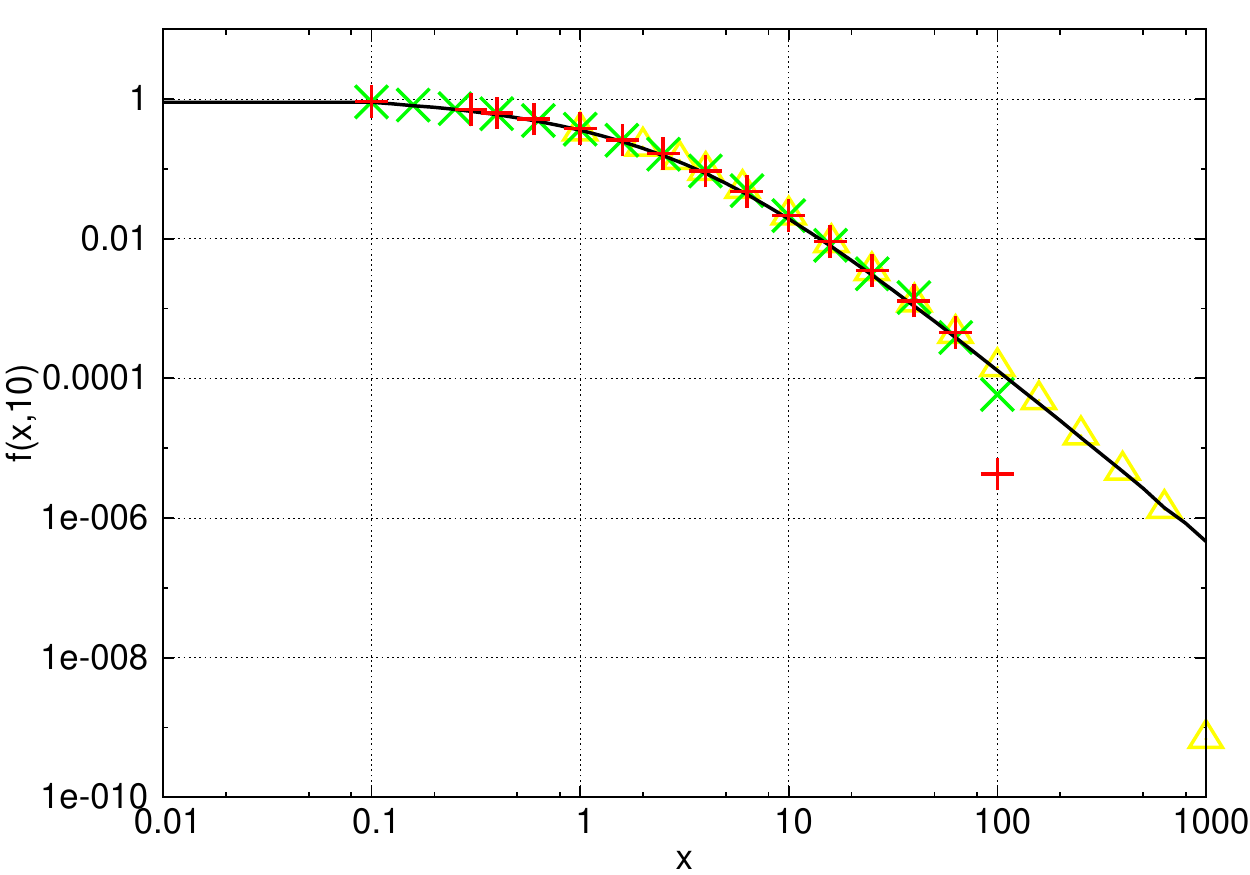}
\label{FIG:FDAES_log}
}
\caption{\small Comparison of numerical methods to the Fourier series solution.
                Black solid line: Fourier series ($N_{max}=10^7, L = 10^5$).
                Yellow triangles: Matrix transfer technique ($\Delta x = 1$).
                Red crosses: Finite-difference method ($\Delta x = 0.1, \Delta t = 0.1$).
                Green crosses: SDE ($n_{particles}=10^5, \Delta t = 0.001$).}
\end{figure}

Second, we compare the different numerical methods to the problem introduced in section 2.2, i.e.\ the space-fractional diffusion-advection equation
\ref{FDAES} with $\kappa=1$, $\alpha=3/2$, and $a=1$. 
The solution to this problem can be approximated by the formula \ref{SOL:FDAES}.\\
From Figure \ref{FIG:FDAES} we can see that the numerical methods compare well with the semi-analytic solution \ref{SOL:FDAES}, to the effect that all methods
reproduce the correct shape of the solution. Note, however, that the solution obtained with the matrix transfer technique has to be corrected by a constant factor,
depending on the resolution of the grid; in this particular case by approximately 0.03. Furthermore, both grid-based methods have, for growing $t$, 
increasing problems at the boundaries, because a finite grid together with a power-law decay of the solution does not fit the given boundary conditions very
 well.\\
In case of the solution derived with the SDE, these problems can be avoided, since again a boundary condition has not to be provided explicitly for this method.
For $t \to \infty$ the solution to this problem evolves into a steady state, which is numerically best approximated by the SDE method, due to the given reasons.

\subsection{Space-fractional diffusion on a finite domain}
\begin{figure}
\subfigure[Introducing boundary conditions in the finite-difference method and for the SDE leads to a deviation from the Fourier series solution in the approximations.
           The matrix transfer technique is in good agreement with the analytical solution.]
{
\includegraphics[scale=0.8]{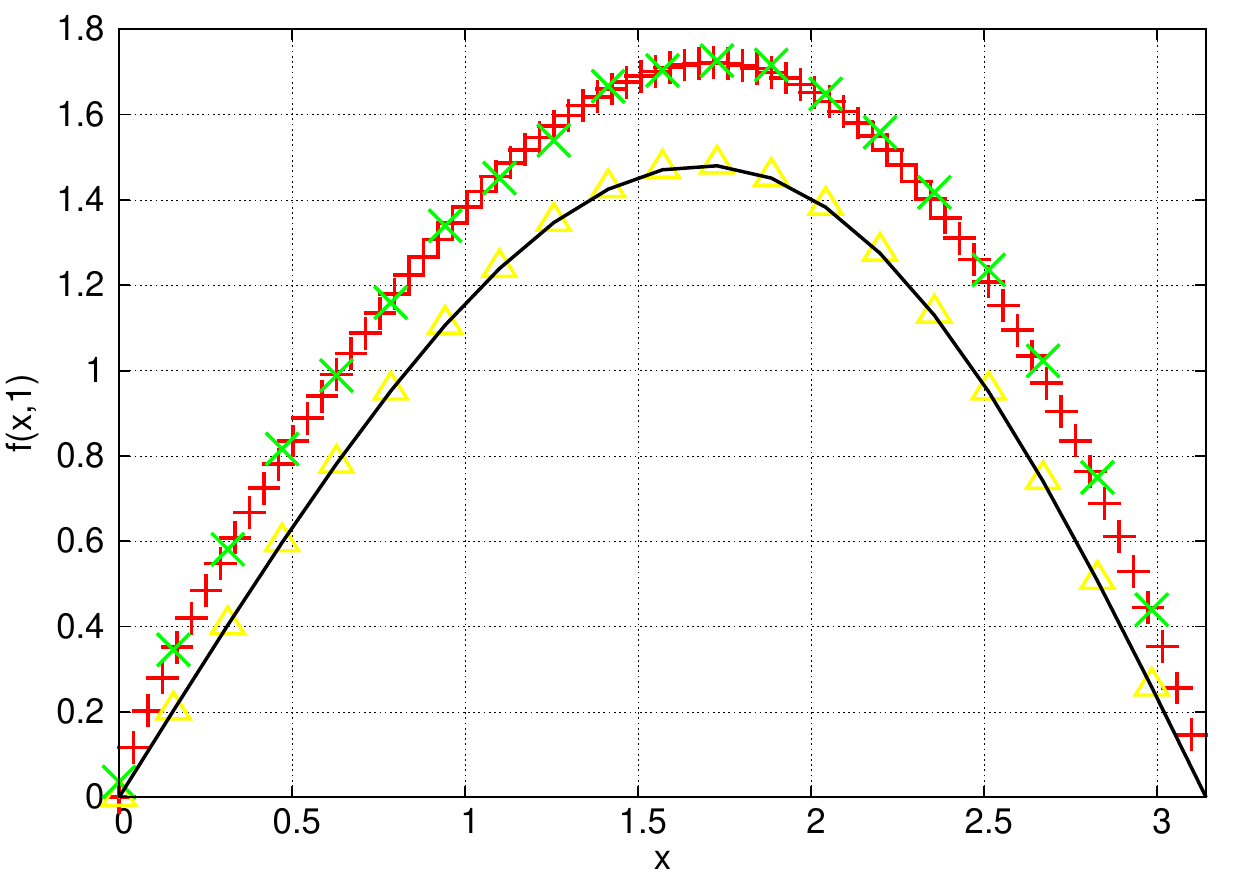}
\label{FIG:ILIC1}
}
\subfigure[Extending the $x$-axis and introducing a periodically continued initial condition leads to a better agreement of the three methods to the Fourier series
           solution.]
{
\includegraphics[scale=0.8]{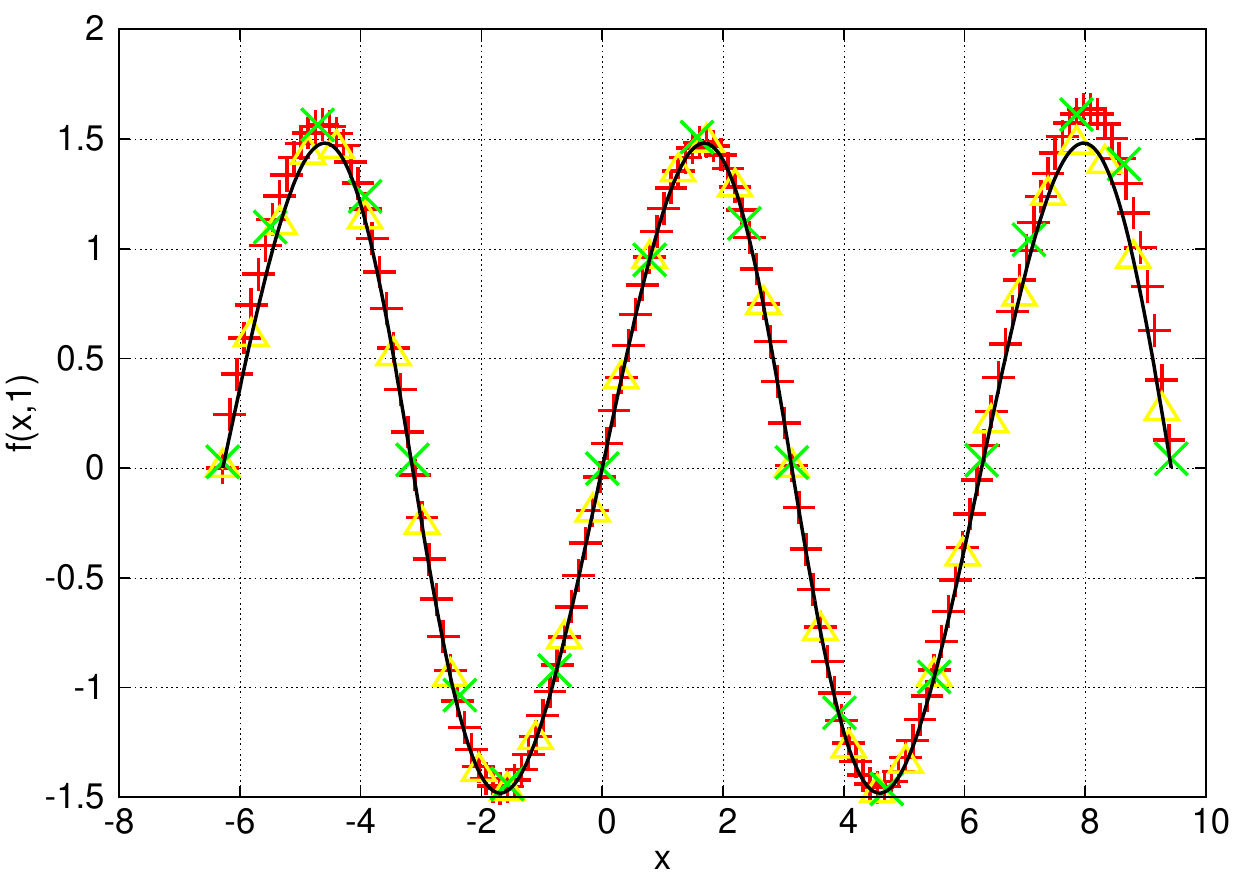}
\label{FIG:ILIC2}
}
\caption{\small{Black solid line: Fourier series solution.
                Yellow triangles: Matrix transfer technique (a: $\Delta x = \pi/20$, b: $\Delta x = \pi/100$).
                Red crosses: Finite-difference method (a: $\Delta x = \pi/300 $, $\Delta t = 1/500$, b: $\Delta x = \pi/800 $, $\Delta t = 1/500$).
                Green crosses: SDE (a,b: $n_{particles} = 10^5$, $\Delta t = 0.001$)}}
\label{FIG:ILIC}
\end{figure}

Finally, we compare the numerical methods to a problem given in Ilic et al.\
\cite{Ilic-etal-2005}, that was used to demonstrate the good performance of the matrix transfer technique. 
This problem is defined by 
\begin{align}
\frac{\partial f}{\partial t} &= \kappa \nabla^{\alpha}f \text{,} \ \ 0 <x <\pi \label{PROB3:1}\\
f(0,t) &= f(\pi,t) = 0 \text{,} \label{PROB3:2} \\
f(x,0) &= x^2(\pi-x) \text{.} \label{PROB3:3}
\end{align}
The solution to this problem was given in \cite{Ilic-etal-2005}, under the assumption that the initial condition can be given in terms of a half-range Fourier sine series as
\begin{equation}
f(x,t) = \sum_{n=1}^{\infty} \left(\frac{8(-1)^{n+1}-4}{n^3} \right) \sin(nx) \exp(-n^{\alpha}\kappa t) \label{PROB3:SOL} \text{.}
\end{equation}
Implicitly, the initial condition becomes periodic with period $L=\pi$, which is of importance as discussed in
the following.\\
In Figure \ref{FIG:ILIC1} we present the solutions obtained with the numerical methods for $\kappa=1$ and $\alpha=3/2$ under the assumptions given in (\ref{PROB3:1}) to (\ref{PROB3:3}). Obviously, the finite-difference method and the SDE solution deviate by the same amount from the exact solution given in (\ref{PROB3:SOL}), whereas the matrix method compares perfectly well.\\
We remark that the finite-difference method was not designed for the solution of the space-fractional diffusion equation with the Riesz operator, which contains an
infinite $x$-axis (see equation \ref{riesz}). This method is, due to the grid, limited to a finite $x$-axis and is in fact solving the space-fractional diffusion equation
\begin{equation}
\frac{\partial f}{\partial t} = \kappa \left(- \frac{_{0}\mathrm{D}^{\alpha}_{x} + _{x}\mathrm{D}^{\alpha}_{\pi}}{2\cos(\frac{\pi \alpha}{2})} \right) f
\end{equation}
and so does the SDE solution, when boundary values are introduced. However, the finite-difference method returns reasonable approximations, even for problems on unbounded domains, if we extend the $x$-axis sufficiently (see previous examples). \\
In Figure \ref{FIG:ILIC2} we demonstrate the effect of an extended $x$-axis for all mentioned numerical methods and the assumption that the initial condition is periodically continued.     In the interval $x \in [0, \pi]$ all numerical methods are now in good agreement with the exact solution. This illustrates, that the exact solution \ref{PROB3:SOL} is only valid using the space-fractional Riesz operator, together with an initial condition given on the whole $x$-axis. In that sense, boundary value problems are problematic if the Riesz operator is involved.

\smallskip
\section{Discussion and conclusions}
The present work provides a critical assessment of numerical solutions of the
space-fractional diffusion-advection equation, which is of high significance
for applications in various natural sciences. In view of the fact that, in 
contrast to the case of normal (Gaussian) diffusion, no standard methods and 
corresponding numerical codes have been established yet, it is 
of importance to critically assess the accuracy and performance of existing 
approaches. 

We have investigated and partly extended three numerical methods, namely a finite-diffe\-rence method
developed by Meerschaert and Tadjeran \cite{Meerschaert-Tadjeran-2006}, the
matrix transfer technique introduced by Ilic et al.\ \cite{Ilic-etal-2005,Ilic-etal-2006},
and the SDE-based Monte-Carlo method employed by Magdziarz and Weron \cite{Magdziarz-Weron-2007},
by applying them to three selected test problems for which analytical or semi-analytical 
solutions were known or have been derived in the present work.

After having demonstrated that a Fourier series representation of the exact solutions to the 
space-fractional diffusion and diffusion-advection equation is, in practice, superior to a 
representation based on Fox's $H$-function, comparisons of the numerical approximations with
the former has revealed: 

\begin{itemize}
\item The finite-difference and the SDE method are suitable for a numerical solution 
      of the space-fractional diffusion-advection equation;
\item The matrix transfer technique appears to be limited to the case of vanishing 
      advection, if grid-dependent correction factors are to be avoided;
\item The finite-difference method is generally computationally expensive as compared 
      to both the matrix transfer technique and the SDE-based Monte-Carlo method,
      particularly depending on the given boundary conditions;
\item The finite-difference method is performing better if
      the Riesz fractional derivative is avoided and instead finite intervals are used
      in a Riemann-Liouville representation;
\item The SDE method appears to have similar problems if non-trivial boundary conditions 
      have to be considered;
\item Extending the interval of interest allows to detour the latter two limitations,
      but obviously makes both methods computationally more expensive.
\end{itemize}

Despite the fact that none of the tested methods is optimal for all space-fractional 
diffusion-advection problems, i.e.\ that all have certain drawbacks, 
we would, based on our numerical studies, give preference to the SDE method that eventually, although
with some effort, could deal successfully with all considered test cases. Beyond the assessment 
made in this paper, this is motivated by the fact that the SDE method is not only attractive in view 
of its rather simple solution algorithm but also from an application-point of view: 
It is often not required to compute a solution everywhere in 
a considered system but only at those locations where measurements have been made. While with the 
finite-difference method and the matrix transfer technique one is forced to compute the solution 
in the whole system, the SDE method is computationally advantageous because its ability to have 
the solution computed for selected locations only. Nonetheless, having several solution methods 
at one's disposal is beneficial, at the least for testing purposes. 

\section*{Acknowledgements}
The authors are grateful to M.M.\ Meerschaert for providing the code ``2D\_CN\_ADI.f90'' and to G.\ 
Zimbardo for extended discussions on anomalous transport. Furthermore, we acknowledge partial support
from the Deutsche Forschungsgemeinschaft (DFG) via the two projects FI~706/8-2 and
FI~706/9-1, and via the DFG Research Unit FOR 1048 ``Instabilities, Turbulence and Transport in Cosmic
Magnetic Fields''. T.S.\ was supported in part by the NSF grant DMS-1108780. 



 \bigskip \smallskip

 \it

 \noindent
$^1$ Institut f\"ur Theoretische Physik IV, 
     Ruhr-Universit\"at Bochum,
     Universit\"atsstrasse 150,
     D-44780 Bochum,
     Germany\\
     e-mail: hf@tp4.rub.de\\[3pt]
$^2$  Interdisciplinary Centre for Advanced Materials Simulation (ICAMS),
      Ruhr-Universit\"at Bochum, Universit\"atsstr.\ 150, 44780 Bochum,
      Germany\\[3pt] 
$^3$ Department of Mathematics, University of Waikato, PB 3105, Hamilton, New Zealand\\[3pt]
$^4$ Department of Mathematics, College of Staten Island, CUNY, NY, USA

\end{document}